\documentclass[
twocolumn,
amsmath,amssymb,
aps,
pra,
 ]{revtex4-1}
\newcommand{\Th}{\mathrm{Th}}

\usepackage[T1]{fontenc}
\usepackage{graphicx}
\usepackage{dcolumn}
\usepackage{bm}
\usepackage{geometry}
\usepackage{color}

\begin{document}

\title{Nuclear clocks based on resonant excitation of ${\gamma}$-transitions}

\author{Ekkehard Peik}
 \email{E-mail address: ekkehard.peik@ptb.de}
\author{Maxim Okhapkin}
 \affiliation{Physikalisch-Technische Bundesanstalt, Bundesallee 100, 38116 Braunschweig, Germany}

\date{12.12.2014}

\begin{abstract}
We review the ideas and concepts for a clock that is based on a radiative transition in the nucleus rather than in the electron shell. This type of clock offers advantages like an insensitivity against field-induced systematic frequency shifts and the opportunity to obtain high stability from interrogating many nuclei in the solid state.
Experimental work concentrates on the low-energy (about 8 eV) isomeric transition in $^{229}$Th. We review the status of the experiments that aim at a direct optical observation of this transition and outline the plans for  high-resolution laser spectroscopy experiments.  
\end{abstract}

\maketitle

\section{Introduction}

Great progress has been made in the development of atomic clocks based on hyperfine and optical transitions of the electronic  structure (see contributions in this special issue and recent reviews \cite{Poli:2013,Ludlow:2014}) and no fundamental limitation to the achievable accuracy of these clocks seems to be in sight. Yet, it may be highly inspiring and instructive to also investigate substantially different approaches in the development of high-performance clocks, like a nuclear clock, as we will discuss here. The term ``nuclear clock'' is sometimes used in the context of methods of radiometric dating, where the age of archeological finds, rocks or meterorites, is determined based on isotope ratios of elements undergoing radioactive decay. These methods should rather be described as a ``stop watch'' on protohistoric or even cosmological time scales. The nuclear clocks that we are going to discuss here are based on the resonant excitation of radiative nuclear (i.e. $\gamma$-ray) transitions, excited by a coherent oscillator. A common feature and important advantage of the two very different types of ``nuclear clocks'' is that the relevant nuclear properties (decay rates and transition frequencies) are hardly influenced by the environment of the nucleus, including the physical or chemical state of the matter. 
           
There are three practical motivations that make the investigation of a nuclear clock attractive and promising:

{\em 1.: Higher accuracy through smaller systematic frequency shifts:} Most of the advanced optical clocks investigated today are limited in accuracy by systematic frequency shifts through external electric or magnetic fields. In clocks that are not operated at cryogenic temperature the light shift induced by thermal radiation from the environment requires careful control and correction. Due to the smallness of the nuclear moments such shifts on a nuclear transition are much smaller, as we will discuss in detail in Section III.

{\em 2.: Higher stability in a solid-state optical clock:} Because of the required precise control of interactions among the atoms and with the environment, the most advanced optical clocks investigated today make use of low atom numbers ranging from one (a single trapped ion) up to a few thousand (in optical lattice clocks) in ultrahigh vacuum. 
The quantum projection noise limited instability (Allan deviation) of a clock is given by 
\begin{equation}
\sigma_y(\tau) =C \frac{\Delta \nu}{\nu_0}\sqrt{\frac{t_c}{N\tau}},
\end{equation}
where $\Delta \nu$ is the linewidth, $\nu_0$ the frequency, $t_c$ the time required for one interrogation cycle, $N$ the number of atoms and 
$C$ a numerical constant of order unity.
In a nuclear clock it may be possible to obtain narrow and unperturbed resonances from a large ensemble of atoms ($N\gg 10^{10}$) in a solid, resulting in a potentially huge gain in signal to noise ratio  and clock stability. M\"ossbauer spectroscopy of nuclear resonances in solids at cryogenic temperature has yielded line Q-factors $\nu_0/ \Delta \nu$ approaching $10^{15}$ \cite{Potzel:1988} and it seems attractive to replace the spontaneously emitted $\gamma$-radiation that is used as the source in these experiments by a coherent oscillator with the qualities of a tunable laser.

{\em 3.: Higher stability through higher reference frequency:} Besides linewidth and signal to noise ratio, the reference frequency $\nu_0$ is the dominant parameter determining the stability of an atomic clock. It is also the one where scaling up promises the most important opportunity for further gains, considering that the typical energy range of M\"ossbauer transitions is tens of keV, i.e. 4 orders of magnitude higher than in present optical clocks.
However, the development of coherent radiation sources and of the required clockwork for the counting of periods
seems to pose major challenges because materials for amplifiers or mixers that provide a similar efficiency as it is available in the visible spectral range are not known. 

Fortunately, Nature has provided a system for the development of a nuclear clock in the frequency range that is accessible with present laser technology: the low-energy transition in $^{229}$Th. This system has largely motivated the conceptual developments that we will discuss here and it is also the object of the experimental projects in this field today.

\section{The low-energy isomer of $^{229}{\rm Th}$}

$^{229}$Th seems to be a unique system in nuclear physics in that it possesses the only known isomer with an excitation energy in the range of optical photon energies and in the range of outer-shell electronic transitions. $^{229}$Th is part of the decay chain of $^{233}$U and undergoes $\alpha$-decay with a half\/life of 7880 years. Its energy level structure has been studied by the group of C. Reich at the Idaho National Engineering Laboratory since the 1970s, mainly relying on spectroscopy of the $\gamma$-radiation emitted after the $\alpha$-decay of $^{233}$U \cite{kroger,helmer}. It was noted that the lower energy levels belong to two rotational bands whose band heads must be very close, one being the ground state, the other the isomer. By evaluating several $\gamma$-decay cascades, the value $(3.5\pm 1.0)$~eV was obtained for the isomer energy \cite{helmer}. Further studies confirmed and extended the knowledge on the overall nuclear level scheme of $^{229}$Th \cite{gulda,barci,ruchowska}. The isomer decays to the ground state under the emission of magnetic dipole radiation with an estimated lifetime of a few 1000~s in an isolated nucleus \cite{ruchowska,dykhne}, providing a natural linewidth in the millihertz range.

The results on the extremely low excitation energy of the isomer inspired a number of theoretical studies, investigating the decay modes of the isomer in different chemical surroundings and possible ways of exciting it with radiation (see Refs. \cite{matinyan,tkalya} for reviews). 
An important concept in this context are electronic bridge processes: Radiative nuclear transition rates may be strongly enhanced if the frequency of the $\gamma$-photon  is resonant with a transition in the electron shell. Similar processes are also discussed under the names bound internal conversion (BIC) or nuclear excitation by electron transition (NEET), where energy is transferred between nuclear and electronic excited states in the near field, accompanied by photon emission (or absorption) through the much stronger electronic transitions.
  
In early experiments, two false detections of the decay of the isomer following $\alpha$-decay of $^{233}$U were reported \cite{irwin,richardson}, but it was quickly clarified that the observed light was luminescence induced by the background of $\alpha$-radiation \cite{utter,shaw}. All further attempts at a direct observation of the optical transition failed. New and better experimental data on the transition energy became available only after a group at LLNL used a cryogenic high-resolution $\gamma$-spectrometer and measured two decay cascades very precisely.  The eventual result for the transition energy is $(7.8\pm0.5)$~eV \cite{beck}, placing the transition in the vacuum-ultraviolet (VUV) at about 160 nm. Radiation at this wavelength is strongly absorbed in air, water and most optical glasses, which explained the failure of  most of the previous attempts to detect the radiation. 
Because of the higher resolution of the  $\gamma$-spectroscopy presented in Ref. \cite{beck}, the new energy value is more reliable than the older result from Ref. \cite{helmer}. Some concerns about the lineshape and a possible blending of unresolved components in the $\gamma$-spectra remain and were taken into account in the uncertainty of $\pm0.5$~eV. An experiment with a magnetic microcalorimeter as a $\gamma$-detector of very high resolution that could determine the energy to an uncertainty of $\approx0.1$~eV  is presently in preparation \cite{Kazakov:2014}.

The expected transition wavelength of about 160~nm is in a range that is accessible by frequency upconversion of laser sources. Thus it appears possible to study this nuclear transition of $^{229}$Th with methods of high-resolution laser spectroscopy.
Promising results on VUV laser spectroscopy  have been obtained with harmonic generation from near-infrared femtosecond lasers \cite{Witte:2005,Gohle:2005}.
In this approach, the frequency of the laser modes can be controlled and measured in the infrared spectral region, while the conversion of the fundamental spectrum  into a sequence of harmonics makes the ensuing measurement precision available in the VUV.
The $^{229m}$Th excitation energy is now known with a relative uncertainty of about 6\%. This big uncertainty presents a major obstacle that has to be overcome  by a more precise measurement before high-resolution laser spectroscopy on the transition can be started. 

Fig. 1 summarizes the relevant spectroscopic information that is known today. The ground state magnetic dipole moment $\mu$ and electric quadrupole moment $Q$ have been determined from the hyperfine structure of transitions in $^{229}$Th$^{3+}$ 
\cite{Safronova:2013}, and the latter also from Coulomb excitation of the nucleus \cite{Bemis:1988}. The nuclear moments of the isomer have been obtained from general considerations and from calculations of the $^{229}$Th nuclear level structure, yielding $\mu\approx -0.08~\mu_N$ \cite{dykhne} and $Q\approx 1.8\times 10^{-28} e$m$^2$ \cite{Tkalya:2011, Litvinova, Campbell:2012}. 

\begin{figure}[h]
\centering
\includegraphics[width=7cm]{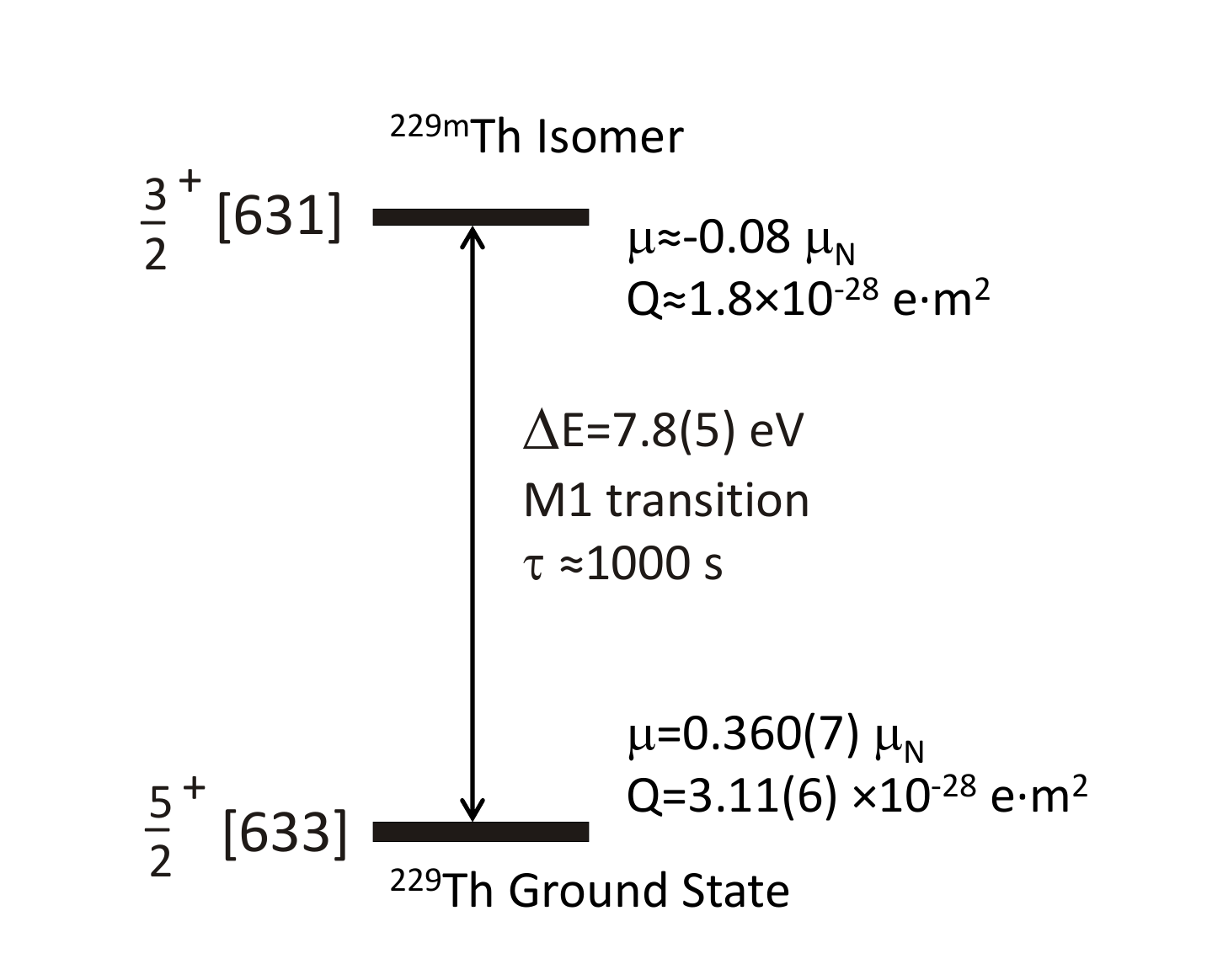}
  \caption{The ground state and lowest excited state of the $^{229}$Th nucleus with the level classification in the Nilsson model. The radiative lifetime 
	for the magnetic dipole transition is taken from Refs. \cite{ruchowska,dykhne}, magnetic moments $\mu$ in nuclear magnetons and quadrupole moments $Q$ from \cite{Safronova:2013, dykhne, Tkalya:2011}.}
\end{figure}

\section{The  $^{229}{\rm Th}$ nuclear clock and its expected performance}

In general, nuclear transition frequencies are less sensitive against external perturbations than transition frequencies of the electron shell because the characteristic nuclear dimensions and nuclear moments are small compared to those of the shell. 
This makes nuclear transitions attractive as highly accurate frequency references with small field-induced shifts \cite{peik2}. 
Apart from motional frequency shifts that can be well controlled e.g. in laser-cooled trapped ions or in a crystal lattice through cryogenic cooling, the interaction with ambient electric or magnetic fields usually is the dominant source of systematic uncertainty in optical frequency standards \cite{Poli:2013,Ludlow:2014}.

If the nuclear transition is not interrogated in a bare nucleus, all estimates on the magnitude of systematic frequency shifts must consider the coupling of the nuclear and electronic energy level systems through the Coulomb and hyperfine interactions. 
It is sometimes argued intuitively that closed electron shells shield external electric fields at the nucleus. However, this holds only for the linear Stark effect in the nonrelativistic limit \cite{Schiff:1963}. The interaction of the nuclear quadrupole moment with external field gradients may actually be enhanced in heavy atoms, an effect called Sternheimer antishielding. For Rn-like Th$^{4+}$ an inversion and enhancement of the field gradient by a factor of $-178$ has been calculated \cite{Feiock:1969}.
Since the effects of external fields depend on both, the electronic and nuclear moments and quantum numbers, it will therefore be important to select a suitable electronic state for the nuclear excitation.   

If we consider for simplicity an $LS$ coupling scheme, the eigenstates of the coupled electronic and nuclear system are characterised by sets of quantum numbers $|\alpha,I;\beta,L,S,J;F,m_F\rangle$, where $I$ denotes the nuclear spin, $L,S,J$ the orbital, spin and total electronic angular momenta, and $F$ and $m_F$ the total atomic angular momentum and its orientation, with the quantization axis defined by the Zeeman effect in a weak external magnetic field.  
$\alpha$ and $\beta$ label the nuclear and electronic configurations. 
In the nuclear transition, the nuclear and total angular momentum quantum numbers ($\alpha, I, F, m_F$) can change, while the purely electronic quantum numbers ($\beta,L,S,J$) remain constant. (We are considering a situation where electronic bridge processes can be neglected because all electronic transitions are far detuned from the narrow nuclear resonance line.) 
In this case, the nuclear transition frequency is independent of all mechanisms that produce level
shifts depending only on the electronic quantum numbers ($\beta,L,S,J$), because these do not change and consequently the upper and the lower state of the transition are affected in the same way. 
To further reduce the sensitivity of the transition frequency to shifts resulting from coupling between electronic and nuclear momenta one may select electronic states with small values of $J<1$ or $F<1$ that eliminate some higher order and tensorial interactions, like the tensor part of the quadratic Stark effect and the interaction between the atomic quadrupole moment and electric field gradients \cite{peik2}. 
An alternative that is applicable for all values of $J$ is the use of stretched states with maximum values of $F$ and $|m_F|$ as proposed in \cite{Campbell:2012}, because these are eigenstates in the coupled as well as in the uncoupled basis: $|F=I+J,~m_F=\pm F\rangle=|J,~m=\pm J\rangle \otimes |I,~m=\pm I\rangle$.

The dependence of the nuclear transition frequency on magnetic fields is quite familiar from atomic clocks based on hyperfine transitions.  
In order to avoid the influence of the linear Zeeman effect, an electronic state can be chosen such that $F$ is an integer, so that a linearly magnetic field-independent   Zeeman component $m_F=0 \rightarrow 0$ is available. In the case of the stretched states with $m_F\ne 0$, the linear Zeeman shift of the transition does not vanish, but can be compensated by interrogating two Zeeman components that are symmetrically shifted like $(m \rightarrow m')$ and $(-m \rightarrow -m')$ and determining the average of both transition frequencies. This method already is succesfully applied in a number of optical clocks based on electronic transitions. 
In both cases, a quadratic Zeeman effect around zero magnetic field remains, that may be as big as $0.1$~Hz$/(\mu{\rm T})^2$, but usually can be controlled very precisely by working at low magnetic field well below the transition from the Zeeman to the Paschen-Back regime, and by monitoring the field strength via the measurement of a linear Zeeman splitting. 

The systematic shifts arising from the interaction with electric fields are more intricate. 
The scalar part of the quadratic Stark effect $\Delta W=-\alpha_SE^2/2$, which leads to the dominant frequency shifts through static electric fields, collisions, and the shift induced by blackbody radiation in electronic transitions, is mainly a $J$-dependent shift that will be common to both levels of the nuclear transition. This part will therefore not lead to a systematic frequency shift in the nuclear clock. The polarizability $\alpha_S$ results from second-order perturbation theory and is proportional to the square of the electric dipole matrix element, divided by the difference in level energies. Going from electronic to nuclear transitions, the nuclear polarizability can therefore be expected to be more than 10 orders of magnitude smaller than those of the electron shell. 
Considering that the quadratic Stark shift in an electronic optical transition is typically of the order of magnitude $\alpha_S/h\approx 10^{-7}$~Hz/(V/m)$^2$, the Stark effect on the isolated nucleus would be negligible. At this point, however, the coupling between electron shell and nucleus becomes important, leading to the so called hyperfine Stark shift, which depends on $F$ and $m_F$. The effect has been studied in microwave atomic clocks where transitions are driven between different $F$-levels of the same ($\beta,L,S,J$) electronic state: For the hyperfine transition in the Cs clock, the Stark coefficient is $\alpha_S/h\approx 2\times 10^{-10}$~Hz/(V/m)$^2$. An additional effect comes from the monopole part of the Coulomb interaction between electronic and nuclear charge distributions, that can be seen as an isomer shift in electronic transitions. Because the nuclear charge distribution changes in the transition, it interacts slightly different with the electron shell and with an external electric field in both levels. For the stretched states of the $^{229}$Th$^{3+}$ ground state $5f\,^2F_{5/2}$, calculations predict a monopole-mediated differential polarizability that is about three times bigger than the hyperfine-mediated effect, leading to a total sensitivity of about $10^{-12}$~Hz/(V/m)$^2$ \cite{Campbell:2012}.  The resulting relative frequency shift induced by blackbody radiation from an environment at 300~K amounts to $1\times 10^{-22}$ for the $^{229}$Th nuclear clock, compared to $1.7\times 10^{-14}$ for the microwave hyperfine transition in a Cs clock.      

From these general considerations (see \cite{peik2,Campbell:2012} for a more detailed discussion) it can be seen that for the radiative nuclear transition
an electronic state can be selected which makes the hyperfine coupled nuclear transition frequency immune 
against field-induced frequency shifts to a level that can not be obtained for an electronic transition.
It is therefore likely that the achievable precision of such a clock will ultimately be limited by the relativistic Doppler shift due to residual motion of the nucleus only.
For a high precision nuclear clock, a single $^{229}$Th$^{3+}$ ion in a Paul trap seems to be especially promising because its electronic level structure is suitable for efficient laser cooling. A sensitive detection of excitation to the isomeric state will be possible using a double resonance scheme, where 
the change of the nuclear spin is observed in the hyperfine structure of a resonance line of the electron shell \cite{peik2}. No electric dipole transitions originate from the  $^2F_{5/2}$  ground state of Th$^{3+}$ in the range of 1.8 -- 15 eV so that electronic bridge processes are not expected to play an important role for the excitation or decay of the isomeric state. A successful project on trapping and laser cooling of Th$^{3+}$ ions has been started at Georgia-Tech in Atlanta \cite{campbell} (see Sect. 4). Alternatively, the method of quantum logic spectroscopy of trapped ions \cite{schmidt} with an auxiliary ion may be applied to other (higher) charge states of $^{229}$Th that do not have easily accessible electronic transitions and can therefore not be laser cooled directly.

In Ref. \cite{peik2} we also introduced the idea of performing laser M\"ossbauer spectroscopy with $^{229}$Th embedded in a crystal or glass. This was later discussed in detail in Refs. \cite{peik3,relle1,Kazakov:2012}. 
The idea of probing the $^{229}$Th nuclear resonance in  a solid receives increased attention, because this is regarded as a relatively simple way to prepare a bigger ensemble of nuclei for optical spectroscopy, even if there will be some line broadening and shifts. Such a sample may be used in a search experiment where the isomer is excited by spectrally filtered synchrotron radiation and the decay is observed by detecting the fluorescence emission. 
The host crystal should be transparent at the nuclear resonance wavelength $\lambda_0$, a criterion that is fulfilled by a number of candidates like the fluorides of the alkaline earths. The crystal could be lightly doped with $^{229}$Th at a concentration of one nucleus per $\lambda_0^3$ in order to avoid near-field coupling and to reduce the effects  of $\alpha$-radiation-induced damage of the crystal lattice. Still, this would allow to do experiments with $10^{11}$ nuclei per mm$^3$ so that direct fluorescence detection of the resonance radiation would be possible even if the resonant scattering rate is only of the order $10^{-4}$/s per nucleus.

The uncertainty budget of such a  $^{229}$Th solid-state nuclear clock will be different from that of a realization with trapped ions. 
A detailed analysis of the expected clock performance was provided in \cite{Kazakov:2012}, where crystal-induced shifts and broadening effects of the nuclear transition for a thorium-doped calcium fluoride crystal were investigated. The nuclear transition in a solid-state lattice environment is not sensitive to recoil or first-order Doppler effects. However, the interaction of thorium nuclei with the lattice environment causes temperature-dependent broadening and shift 
of the nuclear resonance where the line shape will depend on the spectral distribution of the phonons.  
Temperature-dependent shifts arise from the electric monopole shift caused by interactions with electrons, an electric quadrupole shift in the presence of a non-vanishing electric field gradient in the crystal, mixing between different nuclear magnetic sublevels, and the second-order Doppler effect. 
In insulators with high bandgap like fluorides, rather high internal electric fields and field gradients will be found. The electron charge density at the position of the nucleus will lead to the isomer shift $\Delta f_{iso}=Ze^2\rho_0\langle r^2\rangle/(h\epsilon_0)$, where $Ze$ is the nuclear charge, $\rho_0$ the electron density at the nucleus and $\langle r^2\rangle$ the mean squared nuclear charge radius. The contribution of a 7s electron in thorium would shift the nuclear ground state by $\Delta f_{iso}\approx1$~GHz with respect to its energy in a bare nucleus. An electric field gradient will produce a quadrupole shift that may be of comparable magnitude: F$^-$ interstitials that appear for charge compensation in a Th-doped CaF$_2$ crystal produce a field gradient along the principal axis of about $5\times10^{22}$~V/m$^2$ \cite{Kazakov:2012}. Coupling to the ground state quadrupole moment of $^{229}$Th would produce a splitting and shift of levels by  about 1 GHz. The temperature dependence of these shifts may be eliminated if the crystal is cryogenically cooled to well below the Debye temperature, so that the influence of phonons is effectively frozen out. The magnetic dipole interaction with $^{19}$F nuclei was estimated to be the dominant source of decoherence at liquid nitrogen temperature for the thorium-doped CaF$_2$ crystal. A relaxation rate of $\approx150$ Hz for the optical coherence due to interaction with the random magnetic field caused by the surrounding fluorine was calculated.

A fundamental problem of the solid-state nuclear clocks arises from the huge difference between a long relaxation time for population of the isomeric state and a short coherence time between ground and excited states of the thorium nucleus caused by magnetic dipole interaction. The coherence time lies in the millisecond range and limits the linewidths and useful duration of the excitation pulse. However, the decay of the isomeric nuclear state back to the ground state corresponds to the lifetime of the isomeric state which is six orders of magnitude longer. 
The fractional instability of a solid-state $^{229}$Th nuclear clock  at optimized interrogation parameters was estimated to be in a range that should also be  accessible with optical lattice clocks \cite{Kazakov:2012}. Nevertheless, one can assume that the high performance of a solid-state nuclear clock in association with a simple realization will make the system convenient for many applications.  

\section{Experimental search for the  $^{229}{\rm Th}$ nuclear transition}

As outlined above, the known properties of the $^{229}$Th nucleus have already stimulated many ideas and detailed concepts of a nuclear clock  have been developed.
The main obstacle on the way towards experimental studies is that the transition energy is now known from indirect $\gamma$-spectroscopy with an uncertainty  of 0.5~eV, many orders of magnitude higher than the natural linewidth that is expected to be in the mHz range. To reduce the uncertainty of the transition energy a number of different experiments are in progress.

{\em Recoil nuclei:} 
The $^{229}$Th isomer is populated with about 2\% probability in the  $\alpha$-decay of $^{233}$U and early attempts at an optical detection of the isomer decay have looked for light emission from $^{233}$U sources \cite{irwin,richardson}, without success because of a strong background  of radioluminescence \cite{utter,shaw}. This problem can be mitigated by using  $^{229}$Th recoil nuclei: If the  $\alpha$-decay of 
$^{233}$U occurs close to the surface of the sample, the freshly  produced $^{229}\Th$ may be ejected and can be collected on an absorber (such as CaF$_2$ or MgF$_2$, for instance) placed in front of the $^{233}$U. After a period of accumulation, the absorber is moved from the uranium source to a detection chamber and the emitted photons of the isomeric state decay to the ground state are detected by photomultipliers (PMT). A spectral selectivity can be provided by PMT properties or by VUV bandpass filters. Groups at  Physikalisch-Technische Bundesanstalt \cite{Zimmermann:2010}, the Lawrence Livermore National Laboratory \cite{Burke:2010} and the Los Alamos National Laboratory \cite{Zhao:2012}  have performed related investigations. A report about an observation of the deexcitation of the isomeric state in the experiment at LANL was  published \cite{Zhao:2012}. However, it is not clear from this publication that the deexcitation of the isomer was observed \cite{Peik:2013}. Radioluminescence remains a problem in the recoil experiments because daughter products in the $^{233}$U decay chain are also transferred to the absorber. A time-dependent luminescence due to Cherenkov radiation resulting from the $\beta$-decay of $^{213}$Bi (45.59 min) and $^{209}$Pb 
(3.25 h) appears on the timescale that is expected for the $^{229m}\Th$ decay signal \cite{Peik:2013}. In addition, the low spectral resolution of these recoil experiments cannot provide an accurate measurement of the isomer energy. 

A proposal for a strongly improved version of these experiments was recently reported by a group of researchers from the Ludwig-Maximilians-University, Munich and the GSI Helmholtzzentrum in Darmstadt \cite{Wense:2013}:  The proposed technique is based on a selection of thorium recoil nuclei and separation from other daughter products by the use of an ion guide and a quadrupole mass filter. After passing through the mass filter, the Th ions are focussed and collected on a surface and the deexcitation signal of the isomeric state will be detected with an intensified CCD camera and spectral resolution can be provided with a tunable VUV filter. The proposed technique gives a chance to reduce the time-dependent background signal caused by nuclear decays of daughter products.

While fluorescence detection on stopped recoil ions could be used to measure the lifetime of the isomer and the transition energy, important additional information could be obtained in collinear laser spectroscopy of $^{229m}\Th$ atomic or ionic beams or by in-source/in-jet resonance ionization
spectroscopy \cite{Raeder:2011,Sonnenschein:2012}. Precise results have been obtained on isotope shifts and hyperfine structure of several Th isotopes already, and work is ongoing to produce an intense source of  $^{229m}\Th$.

{\em Experiments with solid samples and synchrotron radiation:}
If a sufficiently large ensemble of $^{229}$Th nuclei in the ground state and a intense source of broadband VUV radiation are available, excitation of the isomeric state is possible without requiring precise information on the transition wavelength. For broadband excitation, one may estimate the excitation rate based on the Einstein B coefficient and the spectral density of radiation intensity: 
$R\approx (c^2 P)/(8\pi h \nu_0^3 \tau A \Delta\nu)$, where $P$ is the light power available in bandwidth $\Delta \nu$ over the focus area $A$, $\tau$ is the isomer lifetime and $\nu_0=1.9\times 10^{15}$~Hz the resonant frequency at 7.8~eV. Assuming $\tau=1000~s$, $P=1.3\times 10^{-16}$~W per $\Delta \nu=1$~Hz bandwidth 
(corresponding to a flux of 100 photons/(s\,Hz) at 7.8~eV) and $A=1$~mm$^2$, one arrives at $R=1\times 10^{-10}~1/$s. The chosen parameters are achievable with undulator radiation from an electron storage ring, for example at the Metrology Light Source (MLS) of the PTB \cite{Gottwald:2012}. One may therefore expect the emission of a detectable steady-state fluoresence signal of 1000 photons per second from a sample of $10^{13}$ nuclei prepared in a transparent solid.

While conceptually simple, the main experimental difficulties in this approach seem to be the production of suitable $^{229}$Th-doped crystals and the control of background luminescence that may be induced by the VUV irradiation. Two groups have made significant progress in this direction: A cooperation of the University of California, Los Angeles, and the Los Alamos National Laboratory is investigating laser host crystals like LiCAF \cite{Jackson:2009} and has made preliminary luminescence investigations under excitation with synchrotron radiation \cite{Rellergert:2010, Hehlen:2013}. A group at the University of Vienna is investigating CaF$_2$ host crystals and has obtained detailed information on the crystal structure and electronic level structure of Th$^{4+}$ in CaF$_2$ \cite{Dessovic:2014}. In an alternative approach, our group at PTB has produced and characterized samples with $^{229}$Th adsorbed on a  CaF$_2$ surface \cite{Yamaguchi}.    

An alternative and efficient method for detection of the isomeric transition of thorium nuclei in a crystal has been proposed that is based on coherent forward scattering \cite{Liao:2012}.  Because the signal would be carried in a directional beam, a detector could be placed further away from the crystal, efficiently reducing the problem of luminescence background.

{\em Experiments with trapped ions:}
The experiments based on laser excitation of the isomeric state in trapped thorium ions seem to be the most precise attempts to determine the nuclear transitions.
At the present time the experiments are carried out with singly and triply charged ions in the PTB and Georgia-Tech respectively. Earlier experiments with trapped Th$^+$ isotopes are reported in Ref. \cite{Kaelber:1989}.
Assuming the energy of the transition corresponds to $(7.8\pm{0.5})$ eV, for direct excitation of the transition a tunable radiation source in the range between $149$ nm and $170$ nm has to be used.  Since such widely tunable lasers in the VUV range are not available, laser excitation can be provided via two-photon electronic bridge or NEET processes (see Fig. 2).

\begin{figure}[h]
\centering
\includegraphics[width=6cm]{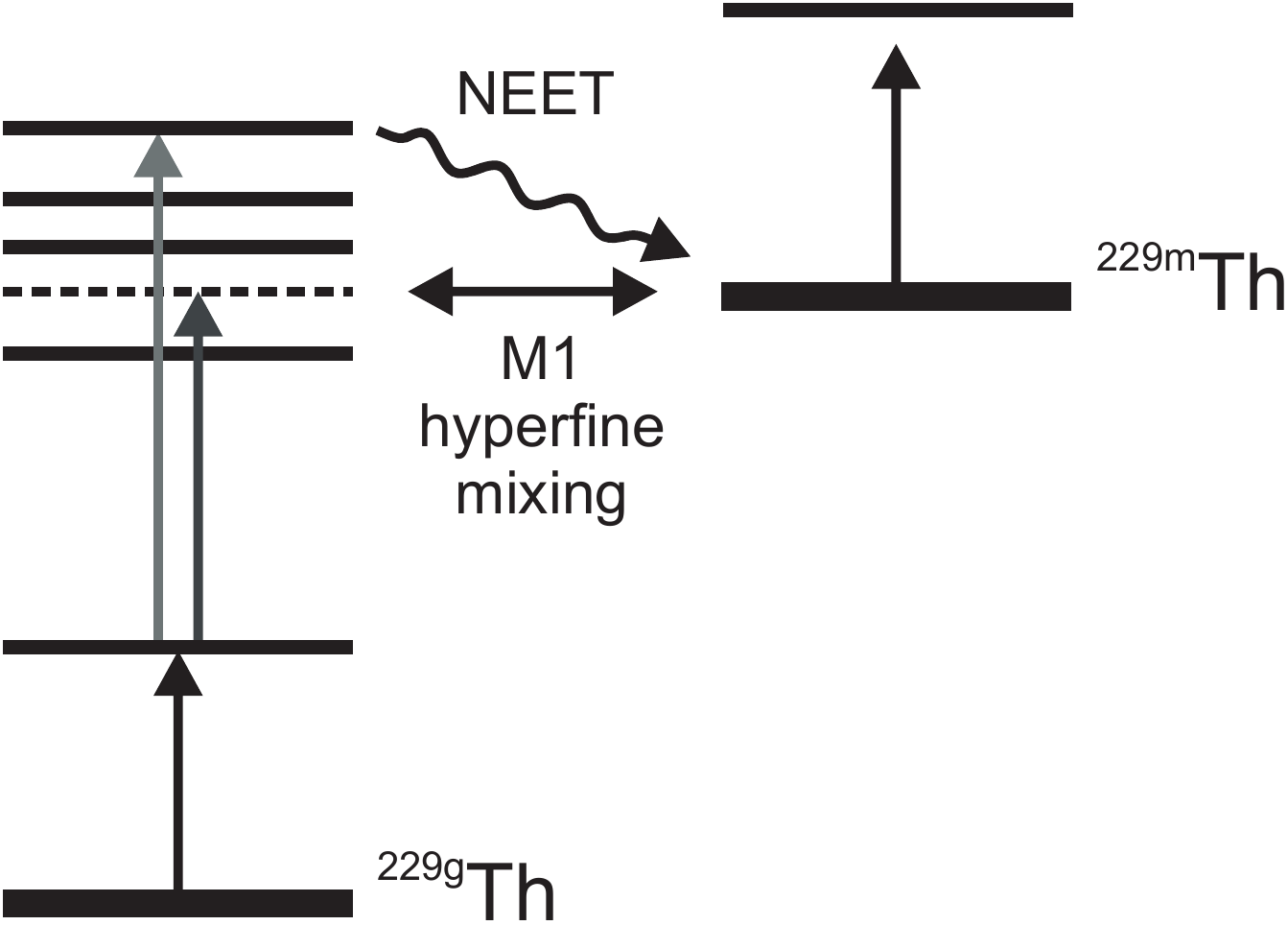}
  \caption{Schematic of the lowest electronic energy levels with nuclear ground state and isomer, showing two-photon excitation of the isomer via a resonant electronic bridge process based on magnetic dipole coupling and via spontaneous emission from an excited electronic state (NEET).}
\end{figure}
   
For our experiments at PTB, we investigate electronic bridge excitation of the isomeric nuclear state of $^{229}\Th^{+}$ using a two-step excitation scheme through the $24874~\mathrm{cm}^{-1}$ intermediate state. Preparatory experiments were carried out with the isotope $^{232}\Th$ which has zero nuclear spin \cite{Herrera:2013}. A linear Paul trap with a capacity of approximately $10^6$ ions of $\Th^{+}$ is loaded by laser ablation from a $\Th$ target \cite{Herrera:2012}.  The ions are cooled to room temperature by Ar buffer gas at $0.1~$Pa pressure. For the first excitation step, pulses from a cw laser diode emitting at $402.0~$nm wavelength are used and the second step is driven by the third harmonic of a tunable pulsed $\mathrm{Ti:Sa}$ laser. 
In scans over the isomeric state energy range from 7.3 to 8.3 eV, 43 previously unknown electronic levels of $\Th^{+}$ were found. The high density of states is close to that  predicted from ab initio atomic structure calculations \cite{porsev2} and makes it likely that an electronic state lies close in energy to the isomer. This will strongly enhance the nuclear excitation rate via the electronic bridge.
At the present time the experiments are carried out with the $^{229}\Th$ isotope and the scan over the search range is in progress. 

In addition to a resonant electronic bridge excitation via magnetic dipole hyperfine mixing described in \cite{porsev2}, the isomer may also be populated via  resonant excitation of higher-lying electronic levels, accompanied by spontaneous emission (NEET, see Fig. 2). 
The efficiency of the isomer excitation via NEET was estimated in Refs.  \cite{Tkalya:1996,Karpeshin:1996, Karpeshin:1999}.
Unfortunately, the low ionization potential of Th$^+$ of about 12~eV causes further resonant three-photon ionization to $\Th^{2+}$ practically for all high-lying states observed in the isomer energy range and reduces the efficiency of NEET excitation. The Th$^+$ signal vanishes within a few minutes at a laser pulse intensity that corresponds to the saturation intensity of the single-photon transitions of the second excitation step.  We assume that doubly ionized thorium is a good candidate for NEET. There are a few known high-lying states in $\Th^{2+}$ with the leading electronic configuration $5f8s$ which can provide an efficient isomer excitation because the wave function of the $s$-electron has the greatest overlap with the nucleus.

$^{229}\Th^{3+}$ possesses a single valence electron outside the Rn-like core and is considered as the most promising system for the development of nuclear clocks with trapped ions. As mentioned before, during the last years  significant progress was established in trapping and laser cooling of Th$^{3+}$ ions. 
Ion production, laser-cooling and the formation of large Coulomb crystals have first been achieved for the long-lived isotope $^{232}$Th$^{3+}$ \cite{campbell}. 
Laser-cooled Wigner crystals of $^{229}\Th^{3+}$ were produced in a linear Paul trap \cite{Campbell:2011}. The hyperfine splitting factors of four levels and the isotopic shifts of three transitions were determined. A value of the nuclear electric quadrupole moment $Q=3.11(6)\times 10^{-28} e$m$^2$ was deduced \cite{Safronova:2013}. The main disadvantage of the triply ionized thorium for a search for the nuclear transition is the presence of only one electronic state ($7p$ at $9.06~$eV) whose energy  is closely above the isomer energy range and can be used for NEET. Alternatively, two-photon resonant electronic bridge excitation from the $7s$ state through the $7p$ level may be an approach for the search \cite{Campbell:2011}. While this electronic bridge excitation scheme has significantly lower enhancement factor than that for $\Th^{+}$, the advantages of an experiment 
with laser-cooled $^{229}\Th^{3+}$ lie in the high spectral resolution and efficient state detection.

An indirect determination of the isomer excitation energy could be possible based on a combination of precision hyperfine
data of the $^{229}\Th^{3+}$ ground state manifolds, accurate atomic structure calculations and the decay rate of the isomer
\cite{Beloy:2014}, where the latter, however, will not be available without a direct observation of the isomer decay.  
 
Experiments with trapped thorium ions were also recently started in the Moscow Engineering-Physics Institute (MEPhI) in collaboration with the Research Institute of Physicotechnical and Radio Measurements (VNIIFTRI) in Russia. An inductively coupled plasma technique for ion loading  from liquid solutions of thorium salts was studied \cite{Troyan:2013} and the development of a quadrupole Paul trap is in progress.

\section{Outlook: Proposals for further studies}

A number of noteworthy effects have been pointed out and studies have been proposed that are related to the coherent manipulation of nuclear states and to applications of a nuclear clock in tests of fundamental principles. 
This shows an interest in this field that originates from several areas of physics and leads to hopes for interesting experimental discoveries, once the initial difficulties have been overcome. 

In Ref. \cite{peik2} we briefly mentioned that the comparison over time of the $^{229}$Th transition frequency with atomic frequency standards opens a new possibility for a laboratory search for temporal variations of the fundamental coupling constants \cite{uzan} as they are predicted by theories of grand unification.  
The idea was developed further by V. Flambaum, who pointed out that the sensitivity of such an experiment would be five orders of magnitude higher than in atomic systems \cite{flambaum1}. This paper raised a controversial discussion (see e.g. Refs. \cite{Hayes, Litvinova}), indicating that the sensitivity to the value of the fine structure constant depends critically on the nuclear structure. It seems likely, though, that the sensitivity will be substantially higher than in presently studied atomic clocks. Flambaum and colleagues finally proposed an experiment that would determine the sensitivity factor by measuring isomeric shifts of electronic transitions in thorium ions \cite{Berengut}. 

Much attention has been attracted by the proposal for a nuclear gamma-ray laser in the optical range, based on stimulated emission of an ensemble of $^{229m}$Th isomeric nuclei in a host dielectric crystal \cite{Tkalya:2011}, an achievement that would meet a long-standing challenge in physics and may open a novel interdisciplinary field of ``quantum nucleonics'' to experimental research.

\bigskip

{\bf Acknowledgements}  

We thank our colleagues at PTB that share with us the interest in the thorium nuclear clock and that have contributed to the work in many discussions and experiments: Chr. Tamm, K. Zimmermann, O. A. Herrera-Sancho, P.  G\l{}owacki, A. Yamaguchi, N. Nemitz, D.-M. Meier, and colleagues from theoretical physics that have helped us with their insights and calculations: V. Flambaum, S. Porsev, V. Dzuba, A. Taichenachev, V. Yudin and M. Safronova.

\end{document}